\documentstyle[11pt]{article}

\begin{document}
\def\beq{\begin{equation}}
\def\eeq{\end{equation}}
\def\beqa{\begin{eqnarray}}
\def\eeqa{\end{eqnarray}}
\begin{titlepage}
\vspace*{-1cm}
\noindent
\phantom{bla}
\hfill{$\scriptstyle{\rm UMHEP-421} \atop{\scriptstyle
{{\rm IPNO/TH95-51}}}$}
\\
\vskip 2.0cm
\begin{center}
{\Large\bf Inverse-Moment Chiral Sum Rules}
\end{center}
\vskip 1.5cm
\begin{center}
{\large Eugene Golowich} \\
\vskip .15cm
Department of Physics and Astronomy \\
University of Massachusetts \\
Amherst MA 01003 USA\\
\vskip .15cm
and \\
\vskip .15cm
{\large Joachim Kambor} \\
\vskip .15cm
Division de Physique Th\'eorique\footnote{Unit\'e de Recherche des
Universit\'es Paris XI et Paris VI associ\'e au CNRS.} \\
Institut de Physique Nucl\'eaire\\
F-91406 Orsay Cedex, France \\
\vskip .3cm
\end{center}
\vskip 1.5cm
\begin{abstract}
\noindent
A general class of inverse-moment
sum rules was previously derived by the authors in a chiral perturbation
theory (ChPT) study at two-loop order of the isospin and hypercharge
vector-current propagators.  Here, we address the evaluation of the
inverse-moment sum rules in terms of existing data and theoretical
constraints.  Two kinds of sum rules are seen to occur, those which contain
as-yet undetermined ${\cal O}(q^6)$ counterterms and those free of such
quantities.  We use the former to obtain phenomenological evaluations of two
${\cal O}(q^6)$ counterterms.  Light is shed on the important but difficult
issue regarding contributions of higher orders in the ChPT expansion.
\end{abstract}
\vfill
\end{titlepage}
\vskip2truecm

\section{\bf Introduction}

Recently, we performed a calculation of the isospin and hypercharge
vector current propagators ($\Delta_{33}^{\mu\nu}(q^2)$
and $\Delta_{88}^{\mu\nu}(q^2)$) to two-loop order in chiral
perturbation theory.$^{\cite{gk},\cite{can}}$  For the sake of reference,
we display the resulting expression for $\Delta_{88}^{\mu\nu}(q^2)$,
\beqa
\lefteqn{\Pi_{88}(q^2 , M_K^2) =
-2L_{10}^{(0)} (\mu^2) -4H_1^{(0)} (\mu^2)} \nonumber \\
& & -6 \left[i{\overline B}_{21}(q^2 ,M_K^2) +
{1\over 192\pi^2} \ln(M_K^2/\mu^2)\right] \nonumber \\
& & + {q^2 \over F_0^2} \bigg[ 18 \big(i{\overline B}_{21}(q^2 ,M_K^2) +
{1\over 192\pi^2} \ln(M_K^2/\mu^2)\big)^2  \nonumber \\
& & - 24 L_9^{(0)} (\mu^2) \big( i{\overline B}_{21}(q^2 ,M_K^2) +
{1\over 192\pi^2} \ln(M_K^2/\mu^2)\big)
- P (\mu^2) \bigg] \nonumber \\
& & - {4M_\pi^2 \over F_0^2} \left[ R (\mu^2)
- {1\over 3} Q (\mu^2) \right] \nonumber \\
&-& {8M_K^2 \over F_0^2} \left[ R (\mu^2) +
{2\over 3} Q (\mu^2) - {3\over 32\pi^2} (L_9^{(0)} (\mu^2)
+ L_{10}^{(0)} (\mu^2) ) \ln(M_K^2/\mu^2) \right] \ .
\label{F7}
\eeqa
$F_0$ is the unrenormalized meson decay constant\footnote{It suffices
to take $F_0 = F_\pi \simeq 0.0933$~GeV, with any error occurring at a
higher order.}, the constants $L_{9}^{(0)} (\mu^2)$, $L_{10}^{(0)}
(\mu^2)$, $H_1^{(0)} (\mu^2)$ are well-known ${\cal O}(q^4)$
counterterms$^{\cite{gl2}}$ and $\mu$ is an arbitrary energy scale. The
function ${\overline B}_{21}(q^2 ,M^2)$, defined in Ref.~\cite{gk}, has
a branch point at $q^2 = 4 M^2$.  Otherwise, it is analytic elsewhere
in the $q^2$ complex plane and in particular, has well-defined
derivatives to all orders at the origin $q^2 = 0$.  The first three terms
on the right hand side of Eq.~(\ref{F7}) are generated at one-loop
order while all the others occur at two-loop order.  Observe the presence
of three constants, $P(\mu^2)$, $Q(\mu^2)$ and $R(\mu^2)$.
These are new ${\cal O}(q^6)$ counterterms which must somehow be determined
from experiment.  It will be the purpose of this paper to thoroughly
analyze `inverse-moment' sum rules which can serve to
phenomenologically constrain the counterterms $P$ and $Q$.
Since the class of such sum rules is of intrinsic theoretical
interest, our study will actually have a more general relevance.

The existence of inverse-moment sum rules can be proved from analyticity
properties of the vector propagators.$^{\cite{kfsr},\cite{kms94}}$  Using
the asymptotic behavior of the functions $\Pi_{aa}(q^2)$ ($a=3,8$ not summed)
implied by the operator product expansion,$^{\cite{svz},\cite{bnp}}$ it
follows that the difference $\Pi_{33}-\Pi_{88}$ satisfies the unsubtracted
dispersion relation,$^{\cite{gk}}$
\begin{equation}
(\Pi_{33}-\Pi_{88})(q^2) = \int_{s_0}^\infty ds \
{(\rho_{33}-\rho_{88})(s) \over s-q^2-i\epsilon} \ \ ,
\label{dr38}
\end{equation}
where the $\{\rho_{aa}\}$ are the corresponding spectral functions.
The real part of this dispersion relation can be rewritten as a set of
sum rules for negative moments of the difference of spectral
functions,
\begin{equation}
\int_{s_0}^\infty  ds \ {(\rho_{33}-\rho_{88})(s)\over
s^{n+1}} = {1\over n!} {{\rm d}^n\over ({\rm d} q^2)^n} (\Pi_{33}-\Pi_{88})(0)
\qquad (n\geq 0) \ \ .
\label{sr1}
\end{equation}
Similar sum rules can be derived for the individual vacuum polarizations.
According to the known asymptotic behaviour, however, at least one
subtraction is required to obtain convergent sum rules.  Thus
we have for $a=3,8$,
\begin{equation}
\int_{s_0}^\infty  ds \ {\rho_{aa}(s) \over s^{n+1}}=
{1\over n!} {{\rm d}^n\over ({\rm d} q^2)^n} \Pi_{aa} (0)
\qquad (n\geq 1) \ .
\label{sr2}
\end{equation}
For definiteness, we shall focus on the quantities
$\Pi_{33}$ and $\Pi_{33} - \Pi_{88}$ in the analysis to follow.

It is convenient to categorize the inverse-moment sum rules
as those which contain ${\cal O}(q^6)$ counterterms and
those which do not.  The former set consists of just two
sum rules, obtained respectively by setting
$n=1$ in Eq.~(\ref{sr2}),
\begin{eqnarray}
\lefteqn{\int_{s_0}^\infty  ds \ {\rho_{33}(s) \over s^2}
= -{1\over F_0^2} P (\mu^2) +
{1\over 480 \pi^2} \left({1\over M_\pi^2}+{1\over 2 M_K^2}\right)}
\nonumber \\
& & + {1\over 8 F_0^2}\left({1\over 16 \pi^2}\right)^2
\left( 1 + {2\over 3}\ln {M_\pi^2\over \mu^2} +
{1\over 3}\ln {M_K^2\over \mu^2} \right)^2 \nonumber \\
& & - {L_9^{(0)}(\mu^2)\over 8 \pi^2 F_0^2}
\left( 1 + {2\over 3}\ln {M_\pi^2\over \mu^2}
+ {1\over 3}\ln {M_K^2\over \mu^2} \right)
\label{Peq}
\end{eqnarray}
and $n=0$ in Eq.~(\ref{sr1}),
\begin{eqnarray}
& & \int_{s_0}^\infty  ds \ {(\rho_{33}-\rho_{88})(s)
\over s} =
{16 (M_K^2-M_\pi^2)\over 3F_0^2} Q (\mu^2)
+ {1\over 48 \pi^2} \ln {M_K^2\over M_\pi^2} \nonumber \\
& & + {M_\pi^2\over F_0^2}\cdot {L_9^{(0)}(\mu^2)+L_{10}^{(0)}(\mu^2)
\over 2 \pi^2} \bigg[ M_\pi^2 \ln {M_\pi^2\over \mu^2} - M_K^2
\ln {M_K^2\over \mu^2} \bigg] \ \ .
\label{Qeq}
\end{eqnarray}
It is from these two relations that we shall obtain determinations
respectively of $P$ and $Q$.

Alternatively, the latter set contains the remaining infinity of
inverse-moment sum rules, with $n\ge 1$ in Eq.~(\ref{sr1}) and $n\ge 2$
in Eq.~(\ref{sr2}).  Each of these sum rules involves only known
quantities on the right-hand side.  Although lacking direct information
on the ${\cal O}(q^6)$ counterterms, such relations will nonetheless
be seen to play a significant role in helping to properly interpret the
inverse-moment sum rules.

\section{\bf Data Analysis}

Sum rules such as those just discussed will be of practical
interest only if one can numerically evaluate the various spectral
integrals which appear.  Because the $s^{-n}$ moments ($n\ge 1$)
strongly emphasize the low energy region, it turns out that the existing
database is sufficient to yield reasonably accurate determinations.
An extensive treatment of the phenomenological extraction of vector and
axialvector spectral functions has been given in Ref.~\cite{dg}.  We
refer the reader to that reference for examples of how
hadronic production data both in $e^+e^-$ annihilation and
in $\tau (1777)$ decay is used to construct spectral functions.

Because the isospin and hypercharge spectral functions $\rho_{33}$ and
$\rho_{88}$ involve isospin-one and isospin-zero channels respectively,
we shall arrange the following discussion accordingly.

\vspace{0.15in}

\begin{center}
{\bf Isospin-one}
\end{center}

\noindent (a) {\it Two-pion}: We adopt the two-pion component of
Ref.~\cite{dg}, except for energies less than $400$~MeV where we
conjoin in a continuous manner the form implied by the two-loop expression
for $\Delta_{33}^{\mu\nu}(q^2)$ obtained in Ref.~\cite{gk}.
In this way, we ensure the proper chiral behaviour near threshold up
to two-loop order.  The agreement between our two-pion spectral function
and low-energy data is displayed in Fig.~1.

\noindent (b) {\it Four-pion}: We employ the form appearing in Ref.~\cite{dg}
without modification.

\noindent (c) {\it Isospin-one $K{\bar K}$}: We employ the approach
given in Ref.~\cite{eid} in which one adopts the SU(3) relation between
pion and kaon form factors to infer the following relation between
the corresponding $e^+e^-$ cross sections,
\beq
\sigma^{(I=1)}_{K{\bar K}}(s) = {\beta^3_{K^-} + \beta^3_{K^0} \over
4\beta^3_{\pi^-} } \sigma_{\pi^+\pi^-}(s) \ \ ,
\label{k1}
\eeq
where $\beta_i \equiv \sqrt{1 - 4M_i^2/s}$.  The resulting extraction
of the $I=1~K{\bar K}$ spectral function is straightforward, and as a
check, is found to yield a branching ratio $B_{\tau\to K^-K^0\nu_\tau}$
in accord with experiment.$^{\cite{eid,pdg,tau}}$

\noindent (d) {\it Asymptotic component}: We employ the form appearing
in Ref.~\cite{dg} without modification.

\vspace{0.15in}

\begin{center}
{\bf Isospin-zero}
\end{center}

\noindent (a) $\omega (782)$:  The state $\omega (782)$ will
contribute to the hypercharge spectral function as a {\it delta function}
rather than as a resonance because we consider only non-anomalous currents
and it would require the anomalous vector current to couple
$\omega (782)$ to the three-pion continuum.  We employ the form
\beq
\rho_{88}^{(\omega)}(s) = F^2_\omega \delta (s - M_\omega^2) \ \ ,
\label{omga}
\eeq
where the constant $F_\omega$ is obtained from the decay rate $\Gamma_{\omega
\to \ell^+\ell^-}$ into lepton-antilepton pairs,$^{\cite{kms94}}$
\beq
F^2_\omega = {9\over 4\pi\alpha^2} M_\omega
\Gamma_{\omega \to \ell^+\ell^-} \ \ ,
\label{fomga}
\eeq
omitting the negligible lepton mass dependence.

\noindent (b) $\phi (1020)$: A compilation of cross section data for
$e^+e^-\to K{\bar K}$ appears in Ref.~\cite{dol}.  It is possible to
infer the $I=0$ $K{\bar K}$ cross section by first subtracting off the
$I=1$ cross section as expressed by Eq.~(\ref{k1}).  The $\phi (1020)$
occurs as a resonance just above the $K{\bar K}$ threshold with full width
$\Gamma_\phi \simeq 4.43$~MeV.  A good fit to the cross section data is
obtained from a relativistic Breit-Wigner resonance form,$^{\cite{pt}}$
\beq
\rho_{88}^{(\phi)}(s)  = {1\over \pi}\cdot
{F^2_\phi M_\phi \Gamma_\phi (s) \over (s - M_\phi^2)^2 +
(M_\phi \Gamma_\phi (s))^2}
\label{phi}
\eeq
where $F_\phi$ is determined by a relation analogous to Eq.~(\ref{fomga})
and
\beq
\Gamma_\phi (s) \equiv {M_\phi\over \sqrt{s}}\cdot
\left[{s - 4 M_K^2 \over M_\phi^2 - 4 M_K^2}\right]^{3/2} \Gamma_\phi \ \ .
\label{gam}
\eeq

\noindent (c) {\it Nonresonant Isospin-zero $K{\bar K}$}:  There is
a small nonresonant component to the $I=0$ $K{\bar K}$ cross
section which is present at higher energies.  Our fit to the
combined resonant and nonresonant components is displayed in
Fig.~2.

\noindent (d) {\it Asymptotic component}:  We follow the recipe
for generating asymptotic form of the $I=1$ spectral
function,$^{\cite{dg}}$ except that the $I=0$ component turns
on at a slightly higher energy, as would be expected from a
study of the contributing multiparticle thresholds.

\section{\bf Analysis}

In our quantitative study of the inverse-moment sum rules, we
consider separately the two classes defined earlier.  First, we analyze
the set of sum rules which do not involve the ${\cal O}(q^6)$
counterterms and then study those that do.  We shall restrict our attention
in this section solely to matters of analysis, leaving questions
of interpretation to Sect.~4.

\begin{center}
{\bf Sum Rules without ${\cal O}(q^6)$ counterterms}
\end{center}

For this class of sum rules, the goal is to determine whether
evaluations of the left-hand sides of the sum rules
agree with those of the right-hand sides.  The results
are studied as a function of the index $n$ which parameterizes the
inverse-moment sum rules ({\it cf.} Eqs.~(\ref{sr1}),(\ref{sr2})).
As described above, existing data is used as input for
numerical evaluation of the spectral integrals which occupy the left-hand
sides of the sum rules.  The right-hand sides are obtained by performing
power series expansions of $\Pi_{aa}(q^2) \ (a=3,8)$ to
yield analytic expressions for the derivative terms,
$({\rm d} /{\rm d} q^2)^n \Pi_{aa} (0) /n!$.  This was done for the
range $n\le 9$ and in addition, numerical studies were carried out for
cases up to $n=20$.  The final step is to focus on the $n$-dependence
of these relations.

A numerical analysis of the sum rules reveals some general patterns:
\begin{enumerate}
\item The two-pion component of the data is by far the most numerically
important, becoming more so as $n$ is increased.
\item Within the two-pion sector, the $\rho (770)$ resonance plays the
major role for $n=0,1$, but threshold values become increasingly important
thereafter.  Some visual insight of this tendency is afforded by
Figure~3, which displays the situation for $n=3$.  Observe how pronounced
the distortion of the two-pion spectral function produced by the inverse
moment has become.
\item For values $n\ge 4$, the $K{\bar K}$ component becomes negligible
due to the large energy of the kaon threshold, $s = 4M_K^2$.
\item Correspondingly, the difference in content between sum rules involving
$\Pi_{33}$ on the one hand and $\Pi_{33}-\Pi_{88}$
on the other becomes negligible for $n\ge 4$.
\end{enumerate}

Our main finding regarding these sum rules
is that they are generally {\it not} satisfied.  Examples
involving $\Pi_{33}$ are shown in Table~1, where differences between
left-hand and right-hand sides of the sum rules are seen to occur quite
generally.  The pattern of discrepancy is, however, far from arbitrary.
Agreement between LHS's and RHS's improves uniformly as $n$ increases.
Indeed, for values $n\ge 6$, the sum rules are obeyed to better than
a few per cent.

\begin{center}
{\bf Sum Rules with ${\cal O}(q^6)$ counterterms}
\end{center}

In the concluding section, we shall provide arguments as to why
the RHS's of the two relations in Eqs.~(\ref{Peq}),(\ref{Qeq}) are good
approximations to the integrals appearing on the LHS's without the need for
implicit higher-order contributions.  That is, for the two sum rules
which contain counterterms, one is justified in neglecting all
contributions of higher order than two-loop in the chiral expansion.

We now turn to an evaluation of the counterterms $P$ and $Q$.
Upon using the phenomenological procedure described in Sect.~2
to perform an evaluation of the spectral integral in Eq.~(\ref{Peq})
we deduce the value
\begin{equation}
P(M_\rho^2)= - (5.6 \pm 0.6) \times 10^{-4} \ \ .
\label{P}
\end{equation}
The choice of scale $\mu = M_\rho$ is a reflection of the important
dynamical role played by $\rho(770)$.  We shall comment more fully
on this point in the concluding section.  The error bar accompanying $P$
is an estimate of the uncertainties associated with our phenomenological
construction of the spectral function $\rho_{33}$ as well as that arising
from the coupling constant $L_9^{(0)}$,
\beqa
L_9^{(0)} ( M_\rho^2) &=& L_9^{(0)} ( M_\eta^2) + {1 \over 128 \pi^2}
\log \left( {M_\eta^2 \over M_\rho^2}\right) \ \ , \nonumber \\
&=& 0.0071 \pm 0.0003 - 0.0005  = 0.0066 \pm 0.0003 \ .
\label{ctL9}
\eeqa

If we adopt the same procedure for counterterm $Q$, we obtain from
Eq.~(\ref{Qeq}) the estimate
\begin{equation}
Q(M_\rho^2) \sim 6.6 \times 10^{-5} \qquad ({\rm Preliminary})\ \ .
\label{Q1}
\end{equation}
We have labeled the above determination `preliminary' because the existing
data sample in the isoscalar channel is incomplete in the four-particle
sector.  Our concern is that, because the counterterm $Q$ is a measure of
$SU(3)$ symmetry breaking, a cancelation between $\rho_{33}$ and $\rho_{88}$
should be evident for each separate $n$-particle sector.  For example, the
resonance sector clearly demonstrates this,
\beq
\int_{s_0}^\infty  ds \ {(\rho_{33}-\rho_{88})(s)\over s}
\bigg|_{\rho,\omega,\phi} = 0.0374 - 0.0103 - 0.0204
\simeq 0.0067\ .
\label{can1}
\eeq
For the four-particle sector, however, one finds
a large value for the purely isovector four-pion
spectral integral ($I_{4\pi} = 0.0107$) which is uncompensated
by a corresponding contribution in the isoscalar channel.  Unfortunately,
there is at present a lack of sufficient data for $e^+e^- \to
K{\bar K}\pi\pi$ to allow an isospin decomposition in that sector.  However,
in order to obtain some measure of the $K{\bar K}\pi\pi$ contribution, we
have treated the $e^+e^- \to K^+{\bar K}^-\pi^+\pi^-$ data of Ref.~\cite{bis}
as if it were purely isoscalar.  Although having an obvious
uncertainty in the magnitude, this approach should get the
threshold and overall energy scale about right.  We obtain
$I_{K{\bar K}2\pi} \simeq 0.0035$ for the associated spectral integral
and finally, the improved determination
\begin{equation}
Q(M_\rho^2) = (3.7 \pm 2.0) \times 10^{-5} \ \ .
\label{Q}
\end{equation}
It is this value that we shall take for our determination of $Q$.
The large error bars indicate the uncertainty in the
four-particle sector.

Often, narrow-width expressions for the $\rho (770)$, $\omega (782)$
and $\phi (1020)$ resonances are used to approximate integrals involving the
physical spectral functions.$^{\cite{kms94},\cite{km}}$  In the narrow-width
approximation, the spectral function for a neutral vector meson of mass
$M_{\rm R}$ is
\beq
\rho (s) = F_R^2 \delta (s - M_{\rm R}^2) \ \ ,
\label{nw1}
\eeq
where the value of $F_R$ is fixed as in Eqs.~(\ref{fomga}),(\ref{gam}).
Omitting any estimates of error bars, we present here for the
sake of comparison the values $P^{({\rm NW})}$ and $Q^{({\rm NW})}$ as
obtained in the narrow-width approximation,
\beq
P^{({\rm NW})}(M_\rho^2) = - 4.0 \times 10^{-4}
\qquad {\rm and} \qquad  Q^{({\rm NW})}(M_\rho^2) =  4.2 \times 10^{-5}  \ \ .
\label{nw2}
\eeq

\section{\bf Conclusions}

It is useful to restate the approach employed thus far
in dispersion-theoretic language.  Thus, consider
the chiral representation $(\Pi_{33} - \Pi_{88})_\chi^{(2)}$ to two-loop
order as obtained in Ref.~\cite{gk} but expressed now as a dispersion
relation.  From known analyticity properties and asymptotic behaviour,
we deduce the once-subtracted form,
\begin{equation}
(\Pi_{33} - \Pi_{88})_\chi^{(2)}(q^2) = a_{38}^{(2)} + q^2
\int_{s_0}^\infty ds \ {(\rho_{33}-\rho_{88})_\chi^{(2)}(s) \over
s(s-q^2-i\epsilon)} \ \ .
\label{chi2}
\end{equation}
In a ChPT framework, the subtraction constant $a^{(2)}_{38}$ corresponds to
an ${\cal O}(q^6)$ counterterm (essentially the quantity $Q$).
Omitting higher orders and simply equating this representation with that
in Eq.~(\ref{dr38}) implies (for $n\ge 0$)
\beq
\int_{s_0}^\infty  ds \ {(\rho_{33}-\rho_{88})(s)\over s^{n+1}} =
a_{38}^{(2)}\delta_{n0} + (1 - \delta_{n0})\int_{s_0}^\infty  ds \
{(\rho_{33}-\rho_{88})_\chi^{(2)}(s)\over s^{n+1}} \ \ .
\label{srchi}
\eeq
This summarizes the content of the sum rules sensitive to $SU(3)$-breaking.

Proceeding in like manner with the $\Pi_{aa\chi}^{(2)}\ (a = 3,8)$, which
would require a twice-subtracted dispersion relation, we have
\beq
\Pi_{aa\chi}^{(2)}(q^2) = a^{(2)}_{aa} + b^{(2)}_{aa}q^2 + q^4
\int_{s_0}^\infty ds \ {\rho_{aa\chi}^{(2)}(s) \over s^2(s-q^2-i\epsilon)} \ \
,
\label{chi3}
\end{equation}
where the $b^{(2)}_{aa}$ are associated with the ${\cal O}(q^6)$
counterterm $P$.  The sum rules of relevance to our calculation are
\beq
\int_{s_0}^\infty  ds \ {\rho_{aa}(s)\over s^{n+1}} =
b^{(2)}_{aa} \delta_{n1} + (1 - \delta_{n1})
\int_{s_0}^\infty ds \ {\rho_{aa\chi}^{(2)}(s) \over s^{n+1}}
\quad (n\ge 1) \ \ ,
\label{chi4}
\eeq
where again we have neglected higher orders.

The above analysis contains several points of interest:
\begin{enumerate}
\item It demonstrates how the ${\cal O}(q^6)$ counterterms of two-loop
ChPT, when viewed in a dispersion relation context, appear as
subtraction constants.
\item It regains the earlier result that our two-loop
ChPT representations lead to an infinity of sum rules, of
which just two constrain ${\cal O}(q^6)$ counterterms.
\item It suggests the following pattern for higher-loop ChPT
treatments --- that ${\cal O}(q^8)$ counterterms from a three-loop
analysis will be constrained by spectral integrals of
$\{\rho_{aa}\}/s^3 \ (a=3,8)$ and of $(\rho_{33} - \rho_{88})/s^2$,
and more generally that ${\cal O}(q^{2n+2})$ counterterms from an $n$-loop
analysis will appear with spectral integrals involving
$\{\rho_{aa}\}/s^{n} \ (a=3,8)$ and $(\rho_{33} - \rho_{88})/s^{n-1}$.
\item Most importantly, it clarifies why the sum rules without
${\cal O}(q^6)$ counterterms must be violated.
\end{enumerate}

The last item deserves comment.  Consider for example the
isospin polarization function $\Pi_{33}$.  The point is that the two-loop
chiral spectral function $\rho_{33\chi}^{(2)}(s)$ is an adequate
approximation to the full physical spectral function $\rho_{33}(s)$
only for energies not too far above the two-pion threshold.  A glance at
Eq.~(\ref{chi4}) shows how a discrepancy must arise in the integrations over
all energy for the sum rules with $n>1$.  The dispersion expressions for
the sum rules also reveal that the discrepancies must decrease as the inverse
moment $n$ increases because the threshold region becomes more and more
enhanced.  As displayed in Fig.~3, the effect is already important for
$n=3$.  Table~2 exhibits the relative importance of the low energy
region lying above the two-pion threshold ($\sqrt{s} \le 0.4$~GeV
to be precise) to that of the full two-pion component.  For $n\ge 6$,
the dominance of the threshold region is almost complete and so the
sum rules are satisfied to a reasonable degree.

Additional physical insight comes from studying the important role played by
the $J=1$, $I=0,1$ resonances, $\rho (770)$, $\omega (782)$ and $\phi(1020)$.
In such resonant channels, dynamical effects of the Goldstone bosons are
generally subordinate since the discontinuities over the two-pion cuts are
too small to compete with a strong resonance.  A well-known analog is the
electromagnetic charge radius of the pion, where the $\rho(770)$ overwhelms
all other effects almost completely.$^{\cite{Leu94}}$  In the ChPT framework,
resonances are not treated as explicit degrees of freedom, but instead
contribute implicitly in the coupling constants of operator counterterms.
Indeed, at ${\cal O}(q^4)$ it was shown$^{\cite{EGPR89}}$ that low-lying
resonances in the vector, axialvector, scalar and pseudoscalar channels
saturate the coupling constants $\{ L_i\} \ (i=1,...,10)$.  Although no such
analysis is currently available beyond one-loop order, we anticipate the same
behaviour --- that a strong resonance will induce large coupling constants for
selected counterterms.$^{\cite{bgs94}}$  Known examples of this type
occur for the processes $\eta\rightarrow \pi\gamma\gamma$$^{\cite{ABBC92}}$
and also in $K\rightarrow \pi\gamma\gamma$$^{\cite{CEP93}}$.

We are thus led to characterize the terms appearing on the RHS's of the
sum rules Eqs.~(\ref{sr1}),(\ref{sr2}) either as `resonant' or as
`continuum'.  A resonant contribution is one which arises from
a counterterm which is saturated by resonance exchange, as in Fig.~(4a).
The continuum contributions are all the others, like those in Fig.~(4b),
which come purely from rescatterings of the Goldstone bosons or
involve a resonance contribution in a subgraph.  As is obvious from
Eq.~(\ref{srchi}) and surrounding discussion, the two-loop sum rules
without ${\cal O}(q^6)$ counterterms have only continuum terms on the RHS.
Such RHS's are recovered entirely from the integrated two-loop
$\rho_{aa \chi}^{(2)} \ (a=3,8)$ spectral functions.\footnote{Although there
may be contributions from resonance dominated counterterms ({\it e.g.}
$L_9^{(0)}$), they nonetheless are continuum contributions arising from
diagrams like Fig.~(4b).}  It is thus clear that sum rules of this type
are not well suited to determine ${\cal O}(q^4)$ counterterms
like $L_9^{(0)}$ which appear only in the continuum part of the RHS.

As an example, we can use a simple vector meson exchange picture to
model the isospin sum rules.  In this vector-dominance (VMD)
picture, we expect the contribution at any given order to reflect the
effect of a counterterm appearing at that order.  Thus, employing the
notation of Ref.~\cite{EGPR89} and working to leading
order, we consider the interaction lagrangian
\begin{equation}
{\cal L}={F_V\over 2 \sqrt{2}} {\rm tr}(V_{\mu\nu} f_+^{\mu\nu}),
\label{GBVint}
\end{equation}
where $V_{\mu\nu}$ is the octet of vector mesons in the antisymmetric
tensor formulation.  The VMD form for the isospin polarization function
is then easily calculated to be
\begin{equation}
\Pi_{33}^{\rm VMD}(q^2) = {F_\rho^2 \over M_\rho^2-q^2} =
\sum_{n=0}^\infty \ {F_\rho^2 \over M_\rho^2} \left(
{q^2 \over M_\rho^2} \right)^n \ \ .
\label{VMDpred}
\end{equation}
The first derivative of this expression evaluated at $q^2 = 0$ leads to a
numerical estimate of the counterterm $P$.  With input parameters
$F_\rho=154$~MeV and $M_\rho=770$~MeV, we obtain
\beq
P^{\rm VMD}(M_\rho^2) = - F_0^2 {F_\rho^2 \over M_\rho^2}
\simeq -5.8 \times 10^{-4} \ ,
\label{wow}
\eeq
which equals, within errors, the phenomenological determination of
Eq.~(\ref{P}).  Note also that within the VMD picture, the natural
scale for counterterm renormalization is $\mu = M_\rho$.  More generally,
we list the derivatives of $\Pi_{33}^{\rm VMD}$ in Table~3, column 2.
We see that for $n\leq 3$ the discrepancy between LHS and RHS
of the sum rules Eq.~(\ref{sr2}) is correctly given by the vector meson
exchange model within $10 \%$ accuracy. For larger $n$, the discrepancy
is small and we expect the corrections to the spectral functions to make up
for the small violation still present.  Of course there will also be mass
corrections to the prediction of Eq.~(\ref{VMDpred}).  However, we have
shown that the large value obtained for the counterterm $P$
as well as the large violation of the sum rules Eq.~(\ref{sr2}) for
small $n$ are well understood in terms of vector meson dominance. There is no
reason for a further large correction of higher order than the order where a
resonance dominated counterterm first appears.

The situation is admittedly not as secure for the $SU(3)$
breaking counterterm $Q$.  It is a fact that the nature of
$SU(3)$ symmetry breaking is still far from completely
understood.   This should serve as a warning that the
dynamics which underlie the value of $Q$ could present a difficult
obstacle.  As we have already commented, progress on the
phenomenological front will be stalled until improved data becomes
available.  As a final thought, we suggest that it might be
profitable to interpret the isovector and isoscalar four-particle
spectral functions in terms of resonances.  Substructure in the four-pion
sector has been cited as evidence for the isovector states
$\rho (1450)$ and $\rho (1700)$.$^{\cite{pdg}}$  To proceed with
this idea, one would need to know more about the resonant couplings
than is presently available.

\begin{center}
{\bf Acknowledgements}
\end{center}

The research described in this paper was supported in part by
the National Science Foundation and by Schweizerischer Nationalfonds.
One of us (J.K.) wishes to acknowledge valuable discussions with
M. Knecht and J. Stern.

\vfill
\eject

\vspace{1.0 in}
\begin{center}
\begin{tabular}{c||ccccc}
\multicolumn{6}{c}{Table~1 {Discrepancy $\Delta (\%)$}}\\
\hline\hline
$n$ & $2$ & $3$ & $4$ & $5$ & $6$ \\ \hline
$\Delta_n (\%)$ & $57.$ & $39.$ & $19.$ & $8.$ & $3.$ \\ \hline
\end{tabular}
\end{center}

\vspace{1.0 in}

\begin{center}
\begin{tabular}{c||cccccc}
\multicolumn{7}{c}{Table~2 Low-energy Contribution to Spectral
Integral} \\
\hline\hline
$n$ & $1$ & $2$ & $3$ & $4$ & $5$ & $6$ \\ \hline
$I_{\rm Low-E}^{2\pi}$/$I_{\rm Total}^{2\pi}$ &
$0.07$ & $0.19$ & $0.44$ & $0.64$ & $0.86$ & $0.94$ \\ \hline
\end{tabular}
\end{center}

\vspace{1.0 in}

\begin{center}
\begin{tabular}{lccc}
\multicolumn{4}{c}{Table~3 {Vector meson contribution to Spectral Integral}} \\
\hline\hline
$n$ & ${1 \over n !} {d^n\over (d q^2)^n} \Pi_{33}^{\rm VMD}(0)$
    & $\int ds {\rho_{33}-\rho_{33\chi}^{(2)} \over s^{n+1}}$
    & $\int ds {\rho_{33} \over s^{n+1}}$  \\ \hline
$1$ & $0.067$ & $0.056$ & $0.083$   \\
$2$ & $0.114$ & $0.103$ & $0.183$   \\
$3$ & $0.192$ & $0.228$ & $0.600$   \\
$4$ & $0.324$ & $0.543$ & $2.944$   \\
$5$ & $0.564$ & $1.45$  & $19.55$   \\
$6$ & $0.921$ & $4.40$  & $154.5$   \\ \hline
\end{tabular}
\end{center}

\vfill \eject

\vspace{0.5 in}
\begin{center}
{\bf\large Figure Captions}
\end{center}
\vspace{0.3cm}
\begin{flushleft}
Fig.~1 \hspace{0.2cm} Low-energy fit to $\rho_{33}(s)$ data.\\
\vspace{0.3cm}
Fig.~2 \hspace{0.2cm} Fit to $I=0$ $K{\bar K}$ cross section. \\
\vspace{0.3cm}
Fig.~3 \hspace{0.2cm} Profile of $\rho_{33}(s)/s^4$. \\
\vspace{0.3cm}
Fig.~4 \hspace{0.2cm} Contributions of (a) resonant and (b) continuum types. \\
\end{flushleft}
\vfill \eject

\end{document}